\documentclass[prd,twocolumn,showpacs,superscriptaddress,nofootinbib,floatfix,showkeys,10pt]{revtex4-1}
\usepackage{graphicx}
\usepackage{amsmath}
\usepackage{bm}
\usepackage{yhmath}
\usepackage{mathtools}
\usepackage{wasysym}
\usepackage[colorlinks,citecolor=violet,urlcolor=violet,linkcolor=blue]{hyperref}
\usepackage{color}
\usepackage{cases}
\usepackage{subfigure}
\usepackage{times}
\usepackage{dcolumn,booktabs,bm}
\usepackage{slashed}
\usepackage{amsfonts,amssymb,stmaryrd,latexsym,amsmath}
\usepackage{textcomp}
\usepackage{multirow}
\usepackage{cancel}
\usepackage{array}
\usepackage{orcidlink}
\usepackage{physics}
\usepackage{siunitx}

\def\SU3{{\text{SU(3)}_{\rm F}}}

\def \pcs4338{{P_{\psi s}^\Lambda(4338)^0}}

\allowdisplaybreaks

\begin{document}
	
	\title{\textcolor{violet}{Fresh look at the nuclear transparency using the generalized parton distributions}}
	
	\author{The MMGPDs\footnote{Modern Multipurpose GPDs} Collaboration:\\
	        Muhammad Goharipour\,\orcidlink{0000-0002-3001-4011}}\email{muhammad.goharipour@ipm.ir}    
    \thanks{Corresponding author}
	\affiliation{School of Physics, Institute for Research in Fundamental Sciences (IPM), P.O. Box  19395-5531, Tehran, Iran}
	\affiliation{School of Particles and Accelerators, Institute for Research in Fundamental Sciences (IPM), P.O. Box 19395-5746, Tehran, Iran}
	
	\author{Fatemeh Irani\,\orcidlink{0009-0008-3447-6361}}\email{f.irani@ut.ac.ir}
	\affiliation{Department of Physics, University of Tehran, North Karegar Avenue, Tehran 14395-547, Iran}
	
	\author{K. Azizi\,\orcidlink{0000-0003-3741-2167}}\email{kazem.azizi@ut.ac.ir} 
	\affiliation{Department of Physics, University of Tehran, North Karegar Avenue, Tehran 14395-547, Iran}
	\affiliation{Department of Physics, Dogus University, Dudullu-\"{U}mraniye, 34775 Istanbul, T\"urkiye}
	\affiliation{School of Particles and Accelerators, Institute for Research in Fundamental Sciences (IPM), P.O. Box 19395-5746, Tehran, Iran}
	             
	\author{\\ And\\ Dipangkar Dutta\,\orcidlink{0000-0002-7103-2849}}\email{d.dutta@msstate.edu}
	\affiliation{Mississippi State University, Mississippi State, Mississippi 39762, USA}	             
	             
\begin{abstract}

 Color transparency (CT) is a fundamental phenomenon in QCD in which hadrons produced in high-energy exclusive processes traverse nuclear matter with minimal interactions. Nuclear transparency, which quantifies this attenuation suppression, is a quantity with high sensitivity to CT effects and provides critical insights into QCD dynamics in nuclear environments.  In this study, we revisit nuclear transparency using the framework of generalized parton distributions (GPDs). By constructing nuclear GPDs (nGPDs) through the incorporation of nuclear parton distribution functions, we calculate the nuclear transparency $ T(Q^2) $ for the carbon nucleus as a function of momentum transfer $ Q^2 $ considering various definitions and compare the results obtained with available experimental data. Our finding highlights the importance of choosing a physically motivated definition of nuclear transparency. Moreover, we emphasize that a more reliable determination of nGPDs requires a dedicated global analysis incorporating nuclear data. Such an approach is essential for improving the theoretical understanding of CT and for achieving consistency with experimental observations in the high-$ Q^2 $ regime.
 
\end{abstract}

%%%%%%%%%%%%%%%%%%%%%%%%%%%%%%%%%%%%%%%%%%%%%%%%%	
	
	\maketitle
	
	\thispagestyle{empty}
	
	\textit{\textbf{\textcolor{violet}{Introduction}}}~~
Color transparency (CT) is an important phenomenon in quantum chromodynamics (QCD) wherein hadrons produced in high-energy (exclusive) processes traverse nuclear matter with minimal interaction~\cite{Jain:1995dd,Dutta:2012ii}. This behavior occurs when the hadron forms a compact, color-neutral configuration, effectively rendering it ``invisible" to the surrounding nuclear medium due to the cancellation of its color fields. As a result, we see a significant suppression in the usual strong interaction between the hadron and the nucleons. This property becomes particularly important in high-momentum-transfer reactions, where small-size color singlet states are favored. The concept is closely tied to nuclear transparency, which quantifies how easily particles can propagate through a nucleus~\cite{Choi:2024gzq}. High nuclear transparency indicates reduced attenuation of the particle as it moves through nuclear matter. So, it can provide a sensitive test for the presence of CT effects. Studies have shown that lighter nuclei typically exhibit higher transparency, and as the energy of the interaction increases, the nucleus is expected to be more transparent. Ongoing and future experiments continue to explore how nuclear transparency varies across energy scales and target materials~\cite{PANDA:2021ozp,Hauenstein:2021zql,Burkert:2022hjz,Choi:2024gzq,CLAS:2025fqh}. Precise measurements of nuclear transparency, principally at higher values of momentum transfer ($Q^2$), provide critical insight into the dynamics of QCD in nuclear environments. Consequently, color transparency and nuclear transparency have always been important subjects to investigate in recent years (see, e.g, Refs.~\cite{Caplow-Munro:2021xwi,Gallmeister:2022gid,Brodsky:2022bum,Miller:2022kxt,Huber:2022wns,Frankfurt:2022cyk,Jain:2022xzo,Das:2023vvn}).

A quantitative interpretation of nuclear transparency in QCD requires a framework that connects the observable to the underlying quark-gluon dynamics. The QCD collinear factorization theorem for certain hard exclusive processes provides such a link~\cite{Diehl:2003ny}. It allows the scattering amplitude to be expressed in terms of nonperturbative generalized parton distributions (GPDs). 
GPDs provide a unified description of the longitudinal‐momentum and transverse‐position correlations of quarks and gluons inside hadrons. Therefore, they enable us to study the hadron structure, particularly the nucleons, in  three dimensions~\cite{Diehl:2003ny,Ji:2004gf,Belitsky:2005qn,Boffi:2007yc,Diehl:2015uka}.
They are experimentally accessed through hard exclusive processes such as deeply virtual Compton scattering (DVCS) and deeply virtual meson production (DVMP) whose factorization properties allow a clean extraction of GPDs over a wide kinematic range \cite{Ji:1996nm,Ji:1998pc,Goeke:2001tz,Belitsky:2001ns,Guidal:2013rya,Kumericki:2016ehc,Mezrag:2022pqk}.
GPDs are also related to familiar hadronic observables, connecting directly to electromagnetic, axial, gravitational, and transition form factors (FFs) \cite{Bernard:2001rs,Guidal:2004nd,Diehl:2013xca,Polyakov:2018zvc,Hashamipour:2020kip,Goharipour:2024atx,Goharipour:2025lep}. It should be noted that GPDs are reduced to ordinary parton distribution functions (PDFs) at the so-called forward limit. When the process involves a nucleus rather than a free proton, medium effects encoded in nuclear PDFs (nPDFs) for the inclusive processes or nuclear GPDs (nGPDs) for the elastic and hard exclusive processes must be taken into account. 

The interplay between the nuclear transparency, color transparency,
and the QCD collinear factorization theorem can be summarized as follows. Nuclear transparency is an experimentally measurable quantity that
reflects the attenuation of a hadron as it propagates through nuclear matter. It serves as a powerful tool for investigating the phenomenon of color transparency. The factorization theorem with GPDs, on the other hand, provides the theoretical tool to compute the relevant exclusive amplitudes, incorporate medium effects, and connect QCD dynamics to the measured transparency.
It was shown that the nuclear transparency can be calculated using GPDs~\cite{Liuti:2004hd,Burkardt:2003mb}. Hence, the precise measurements of nuclear transparency can put new constraints on GPDs and their nuclear modifications. According to QCD predictions, nuclear transparency should exhibit a characteristic rise with increasing $Q^2$ due to the CT phenomenon.  However, recent high-precision measurements of the nuclear transparency of protons~\cite{HallC:2020ijh} challenge this expectation. Actually, the measured nuclear transparency of carbon has been found to be both energy and $Q^2$ independent up to $Q^2 =$ 14.2\,(GeV/c)$^2$. In the present study, we revisit the nuclear transparency phenomenon through the lens of GPDs, aiming to reconcile these experimental findings with theoretical expectations. Our investigation provides new insights into the interplay between GPDs and CT in the high-$Q^2$ regime.

%%%%%%%%%%%%%%%%%%%%%%%%%%%%%%%%%%%%%%%%%%%%%
	
	\textbf{\textit{\textcolor{violet}{Nuclear Transparency}}}~~
The phenomenon of color transparency is a specific prediction of QCD~\cite{Dutta:2012ii} which states that hadrons (like protons or pions) produced in high-momentum-transfer reactions may interact weakly with the nuclear medium while transiting through it. Based on this concept, many experimental efforts haven been performed to measure the nuclear transparency~\cite{HallC:2020ijh,Makins:1994mm,ONeill:1994znv,Abbott:1997bc,Garrow:2001di,Carroll:1988rp,Mardor:1998zf,Leksanov:2001ui,Aclander:2004zm} which refers to the degree to which a nucleus allows particles to pass through it. Such measurements help describe how ``transparent" a nucleus appears to incoming particles, particularly in high-energy collisions. If the nucleus is highly transparent, the particles exit with little energy loss or deflection. It is well established now that transparency is influenced by nuclear density; light nuclei (like carbon) are more transparent than heavy ones (like gold)~\cite{Dutta:2012ii}. Also, at very high energies (e.g., in deep inelastic scattering), the nucleus becomes completely transparent.

From an experimental point of view, nuclear transparency can be measured by studying how particles (such as protons, pions, or electrons) interact with nuclei in high-energy collisions. One of the reactions used to measure nuclear transparency is the quasielastic scattering where a high-energy electron ($ e $) is fired at a nucleus with mass number $ A $ such as those performed at Jefferson Lab (JLab) and SLAC~\cite{HallC:2020ijh}. Due to the scattering, a nucleus with mass number $ A-1 $ is produced in the final state by knocking out a proton ($ p $) from the parent nucleus, $ e + A \rightarrow e^{\prime} + p + (A-1) $. The nuclear transparency can be measured by detecting the scattered electron ($ e^{\prime} $) and ejected proton. To be more precise, the nuclear transparency $ T(Q^2) $, where $ Q $ is the 4-momentum transfer, is defined as the ratio of experimental yield  ($Y_{\text{exp}}(E_m,\vec{p}_m)$) to the expected theoretical yield for a noninteracting proton ($Y_{\text{theo}}(E_m,\vec{p}_m)$)~\cite{HallC:2020ijh}, such as those calculated assuming the plane-wave impulse approximation~\cite{Meier-Hajduk:1983pii}: 
\begin{equation}
\label{Eq1}
    T(Q^2) = \frac{\int_{V} d^3p_{m}dE_{m} Y_{\text{exp}}(E_m,\vec{p}_m)}{\int_{V} d^3p_{m}dE_{m} Y_{\text{theo}}(E_m,\vec{p}_m)}\,.
\end{equation}
Note that $Y_{\text{exp}} $ and $Y_{\text{theo}} $ are integrated over the same phase space volume $ V $. Here, $E_m = \nu - T_{p'} - T_{A-1}$ and $\vec{p}_{m} = \vec{p}_{p'} - \vec{q}$ are the missing energy and momentum, respectively, where $T_{p'}$ and $\vec{p}_{p'}$ represents the kinetic energy  and momentum of knocked out protons, and $T_{A-1}$ is the reconstructed kinetic energy of the $A-1$ recoiling nucleus. As usual, $ \nu $ and $ \vec{q} $ are the energy and momentum transfer, respectively, which can be determined using the electron beam energy or momentum of the incoming electron ($E_e/\vec{p}_e$) and the scattered electron ($E_{e'}/\vec{p}_{e'}$) as $\nu = E_e - E_{e'}$ and $\vec{q} = \vec{p}_{e} - \vec{p}_{e'}$.

As a result, if $ T\approx 1 $, the nucleus is nearly transparent while if $ T\ll 1 $, the proton is strongly absorbed. Note that the nuclear transparency
can also be measured through the hadronic collisions~\cite{Frankfurt:1993it,Frankfurt:1988nt,Ralston:1988rb,Frankfurt:2005mc,Horowitz:2011gd,Lappi:2013am,Kordell:2016njg,Larionov:2025equ} such as those performed at the LHC. By comparing yields from light (e.g., carbon) versus heavy (e.g., lead) nuclei, one can measure the transparency. If protons exit with minimal energy loss, the nucleus is more transparent.

%%%%%%%%%%%%%%%%%%%%%%%%%%%%%%%%%%%%%%	
	
	\textbf{\textit{\textcolor{violet}{GPDs at zero skewness}}}~~ 
It is well established now that GPDs can provide a powerful framework for probing the three-dimensional structure of hadrons, especially nucleons~\cite{Diehl:2003ny,Ji:2004gf,Belitsky:2005qn,Boffi:2007yc,Diehl:2015uka}. They can be accessed through hard exclusive processes such as DVCS and DVMP~\cite{Ji:1996nm,Ji:1998pc,Goeke:2001tz,Belitsky:2001ns,Guidal:2013rya,Kumericki:2016ehc,Mezrag:2022pqk}. In contrary to the ordinary PDFs, which depend only on the longitudinal momentum fraction $ x $ and the factorization scale $ \mu $, GPDs depend also to the negative momentum transverse squared, $ t=-Q^2 $, and the longitudinal momentum transverse (skewness parameter), $ \xi $. GPDs at zero skewness are also an essential ingredient of different types of elastic scattering processes because they are related to different types of hadron FFs including the
electromagnetic, axial, gravitational, and transition FFs~\cite{Bernard:2001rs,Guidal:2004nd,Diehl:2013xca,Polyakov:2018zvc,Hashamipour:2020kip,Goharipour:2024atx,Goharipour:2025lep}.

The unpolarized valence GPDs $H_v^q(x,t)$ and $E_v^q(x,t)$ at zero skewness ($\xi=0$) for quark flavors $q=u,d$ (neglecting the strange-quark contribution) can be determined through a global analysis of the experimental data of the electromagnetic FFs and elastic electron-proton cross sections, as demonstrated by the MMGPDs Collaboration~\cite{Goharipour:2024atx,Goharipour:2024mbk}.  
This approach utilizes two fundamental relations: first,  the connection between nucleon FFs and GPDs which is established through the following sum rules~\cite{Diehl:2004cx,Diehl:2013xca}:
\begin{align}
F_1(Q^2)=\sum_q e_q F^q_1(Q^2)=\sum_q e_q \int_{0}^1 dx\, H_v^q(x,\mu^2,Q^2)\,, \nonumber \\ 
F_2(Q^2)=\sum_q e_q F^q_2(Q^2)=\sum_q e_q \int_{0}^1 dx\, E_v^q(x,\mu^2,Q^2)\,,
\label{Eq2}
\end{align}
where $F_1$ and $F_2$ are the Dirac and Pauli form factors, respectively, and  $ e_q $ refers to the electric charge of the constituent quark $ q $. Note that the relations in Eq.~(\ref{Eq2}) between the Dirac and Pauli FFs and the GPDs
hold for any $ \xi $, but after integration over $ x $, the $ \xi $ dependence drops out due to the polynomiality property~\cite{Diehl:2013xca}.
Moreover, the neutron ($ n $) FFs can be obtained by considering the isospin symmetry $ u^p=d^n, d^p=u^n $. Second, the relation between these FFs and the Sachs electromagnetic FFs is as follows:
\begin{align}
\label{Eq3}
G_M(Q^2) &= F_1(Q^2) + F_2(Q^2)\,, \nonumber \\ 
G_E(Q^2) &= F_1(Q^2) - \frac{Q^2}{4m^2} F_2(Q^2)\,.
\end{align}
By employing these relations within a $\chi^2$ minimization framework, and adopting suitable parametrizations for $H$ and $E$, one can extract the first moments of the GPDs at zero skewness from experimental data on $G_E(Q^2)$, $G_M(Q^2)$, and differential cross sections (or reduced cross sections).

Building on the MMGPDs Collaboration's framework~\cite{Goharipour:2024atx,Goharipour:2024mbk}, the valence quark GPDs are parametrized through the following ansatz:
\begin{align}
H_v^q(x,\mu^2,t) &= q_v(x,\mu^2)\exp\left[t f_v^q(x)\right], \nonumber \\ 
E_v^q(x,\mu^2,t) &= e_v^q(x,\mu^2)\exp\left[t g_v^q(x)\right],
\label{Eq4}
\end{align}
where $q_v(x,\mu^2)$ represents the unpolarized valence quark PDF and the profile functions $f_v^q(x)$ and $g_v^q(x)$ follow the general form:
\begin{equation}
\label{Eq5}
\mathcal{F}(x) = \alpha^{\prime}(1-x)^3\log\frac{1}{x} + B(1-x)^3 + A x(1-x)^2.
\end{equation}
This parametrizations successfully describes the experimental data while naturally reducing to standard PDFs in the forward limit ($t=0$, $\xi=0$).

For GPD $H$, the MMGPDs Collaboration employed next-to-leading order \texttt{NNPDF} valence quark distributions~\cite{NNPDF:2021njg} at $\mu = 2$ GeV as the forward limit. However, since no experimental parametrization exists for the forward limit $e_v^q(x)$ of GPD $E$, the collaboration determined it through a global fit using the functional form:
\begin{equation}
\label{Eq6}
e_v^q(x) = \kappa_q N_q x^{-\alpha_q} (1-x)^{\beta_q} (1 + \gamma_q\sqrt{x}),
\end{equation}
where $\kappa_q$ ($q = u,d$) are the quark magnetic moments, determined by the measured nucleon magnetic moments. It should also be noted that the normalization factor $ N_q $ is obtained from the following sum rule:
\begin{equation}
\label{Eq7}
\int_0^1 dx e_v^q(x)=\kappa_q.
\end{equation}

In high-energy processes involving nuclear targets, the partonic structure must be described by nPDFs rather than free proton PDFs to obtain accurate theoretical predictions~\cite{Goharipour:2017uic,Francener:2025tyh}. These nPDFs characterize the modified quark and gluon distributions within bound nucleons and are determined through global analyses of nuclear deep inelastic scattering and other hard processes~\cite{deFlorian:2011fp,Kovarik:2015cma,Eskola:2016oht,Wang:2016mzo,Khanpour:2020zyu,Eskola:2021nhw,Duwentaster:2022kpv,AbdulKhalek:2022fyi,AbdulKhalek:2020yuc,Helenius:2021tof,Flore:2025uhv,Yang:2025bub} (see Ref.~\cite{Klasen:2023uqj} for a recent review). These nuclear modifications extend to elastic scattering or hard exclusive processes involving nuclei, where nGPDs must replace the free nucleon GPDs. While the most rigorous approach to determine nGPDs at zero skewness is through a direct analysis of elastic electron-nucleus scattering data analogous to Refs.~\cite{Goharipour:2024atx,Goharipour:2024mbk} for the proton GPDs, current limitations in nuclear data availability motivate alternative strategies. Here, we explore the feasibility of constructing nGPDs by appropriately modifying free proton GPDs.

Following the classification in Appendix A of Ref.~\cite{AbdulKhalek:2022fyi}, we identify two distinct types of nPDFs:
\begin{enumerate}
    \item Bound nucleon nPDFs: the distributions $f^{p/A}$ and $f^{n/A}$ describe bound protons and neutrons in a nucleus with mass number $A$, related through isospin symmetry ($u^{p/A} = d^{n/A}$, etc.).
    
    \item Average nucleon nPDF: the nuclear-averaged distribution
    \begin{equation}
        f^{N/A}(x,\mu^2) = \frac{Z}{A}f^{p/A}(x,\mu^2) + \frac{A-Z}{A}f^{n/A}(x,\mu^2)\,,
        \label{Eq8}
    \end{equation}
    represents an isospin-weighted combination for a nucleus with charge $Z$. This normalization is often used when comparing to free-nucleon PDFs.
\end{enumerate}
We correspondingly define two analogous types of nGPDs:
\begin{enumerate}
    \item Bound nucleon nGPDs: using the profile function $\mathcal{F}(x)$ from Eq.~(\ref{Eq5}),
    \begin{align}
        F^{p/A}(x,\mu^2,t) &= f^{p/A}(x,\mu^2)\exp[t\mathcal{F}^{p/A}(x)]\,, \nonumber \\
        F^{n/A}(x,\mu^2,t) &= f^{n/A}(x,\mu^2)\exp[t\mathcal{F}^{n/A}(x)]\,,
        \label{Eq9}
    \end{align}
    with isospin symmetry assumed for both distributions and profile functions. 
In principle, the profile function of a bound proton GPD, $\mathcal{F}^{p/A}(x)$, should differ from the free-nucleon one, $f_v^q(x)$, in Eq.~(\ref{Eq5}), because the nuclear medium can modify the spatial distribution of partons. While its functional form can be taken the same as in Eq.~(\ref{Eq5}), its parameters should, in general, be $A$ dependent. In the present work, due to the lack of precise information on the $A$ dependence of these parameters, we assume the profile functions are universal for different nuclei and identical to the free-proton case. Nevertheless, we study the impact of nuclear effects by considering a simple overall scaling, $\mathcal{F}^{p/A}(x)\propto 1/A^\alpha$, for two 
nominal values of $\alpha$.
    
    \item Average nucleon nGPD:
    \begin{equation}
        F^{N/A}(x,\mu^2,t) = \frac{Z}{A}F^{p/A}(x,\mu^2,t) + \frac{A-Z}{A}F^{n/A}(x,\mu^2,t)\,,
        \label{Eq10}
    \end{equation}
\end{enumerate}
which represents the average GPD per nucleon in the nucleus. This form is natural when comparing to free nucleon GPDs or when normalizing observables on a per-nucleon basis. While these two definitions are studied for completeness, our results show that using the average nucleon nGPD $ F^{N/A} $ provides the most accurate and physically consistent description of the nuclear transparency data (see Fig.~\ref{fig:TSig}). The other definition is explored only to illustrate its limitation and to justify our final choice.

Within this framework, one can employ the nPDFs from the \texttt{nNNPDF} Collaboration~\cite{AbdulKhalek:2022fyi} at $\mu=2$ GeV, maintaining consistency with the MMGPDs Collaboration's use of free-proton \texttt{NNPDF} PDFs~\cite{NNPDF:2021njg} at the same scale for constructing proton GPDs. While this approach easily facilitates the construction of nuclear GPDs $H$, a significant challenge arises for GPDs $E$: currently, no nuclear modifications exist for their forward limits $e_v^q(x,\mu^2)$ from high-energy experimental data. To address this limitation in calculations requiring GPDs $E$, one can consider the following scenarios:
\begin{enumerate}
    \item No nuclear modification: proceed with calculations using only nuclear-modified valence PDFs $q_v(x,\mu^2)$, while keeping $e_v^q(x,\mu^2)$ unmodified from their free-nucleon forms. 
    
    \item Universal nuclear modification: sssume identical nuclear modification factors for $e_v^q(x,\mu^2)$ as those determined for the corresponding $q_v(x,\mu^2)$.
\end{enumerate}

%%%%%%%%%%%%%%%%%%%%%%%%%%%%%%	
	
	\textit{\textbf{\textcolor{violet}{Nuclear transparency in GPDs language}}}~~
	As discussed earlier, several studies have explored calculating nuclear transparency $T(Q^2)$ directly from GPDs~\cite{Liuti:2004hd,Burkardt:2003mb}, typically defined as the ratio of nuclear to nucleon GPD integrals. While this approach provides valuable theoretical insight, we argue that a more robust determination of $T(Q^2)$ can be obtained by considering $ T(Q^2) $ as the ratio of nuclear to nucleon differential cross-section $ d\sigma/dQ^2 $ or simply the reduced cross-section $ \sigma_R $ of the elastic electron-proton scattering. The reason is that the cross-section (whether differential or reduced) inherently incorporates contributions from both $H$ and $E$ GPDs, whereas the GPD-integral approach requires an ad hoc choice between these components (with Refs.~\cite{Liuti:2004hd,Burkardt:2003mb} selecting only $H$).
Moreover, the cross-section is a quantity that is more relevant to the measured yields in Eq.~(\ref{Eq1}) conceptually. This approach thus avoids the theoretical ambiguity in GPD selection while maintaining closer ties to experimental measurements.

According to the framework established in the previous section, we examine alternative approaches for calculating nuclear transparency $T(Q^2)$ using various nGPDs. We consider both GPD-based definitions and those tied directly to experimental observables such as FFs and cross-sections. Following Ref.~\cite{Liuti:2004hd}, we first investigate a GPD-based definition:
\begin{equation}
T(Q^2) = \frac{\left[\int_0^1 dx\, H^A(x,t)\right]^2}{\left[\int_0^1 dx\, H(x,t)\right]^2},
\label{Eq11}
\end{equation}
where $H = H_v^u + H_v^d$ includes only valence quark contributions from GPD $H$ (GPD $E$ is ignored). Note that the cross-section of the (quasi)elastic electron-proton scattering is proportional to the electromagnetic FFs which are related only to the valence quark distributions. Another point should be mentioned is that Eq.~(\ref{Eq11}) has been proposed as a phenomenological ansatz for nuclear transparency in terms of GPDs. The underlying idea is that the integral of $ H(x,t) $ over $ x $ gives the Dirac form factor $ F_1(t) $, and the squared form factor is proportional to the nucleon coherent elastic cross section. In this picture, the ratio of squared integrals for the nucleus and the free nucleon was taken as a proxy for the nuclear transparency. In fact, Eq.~(\ref{Eq11}) is introduced as a phenomenological starting point, inspired by Ref.~\cite{Liuti:2004hd}. Its purpose is to explore the possible role of nuclear GPDs in describing nuclear transparency in a simple, model-independent way, before introducing more sophisticated treatments involving physical observables. It should also be emphasized that the GPD framework is applied to bound nucleons rather
than to the entire nucleus, and that Eq.~(\ref{Eq1}) does not describe coherent nuclear scattering.

\begin{figure}[!htb]
    \centering
\includegraphics[scale=0.73]{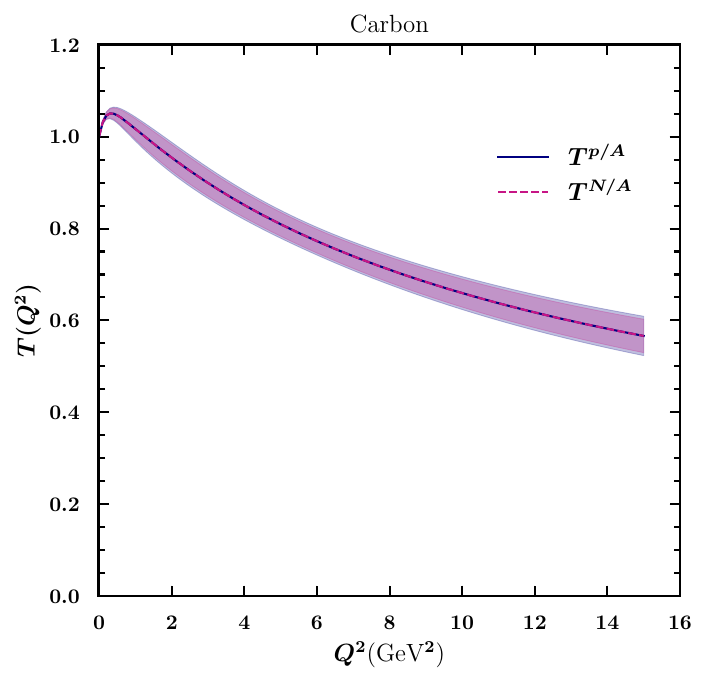}   
    \caption{A comparison between the results of $ T(Q^2) $ for the Carbon nucleus obtained using $ H^{p/A} $ and $ H^{N/A} $ nGPDs, denoted as $ T^{p/A} $ and $ T^{N/A} $, respectively.}
\label{fig:TH}
\end{figure}

\begin{figure}[!htb]
    \centering
    \includegraphics[scale=0.73]{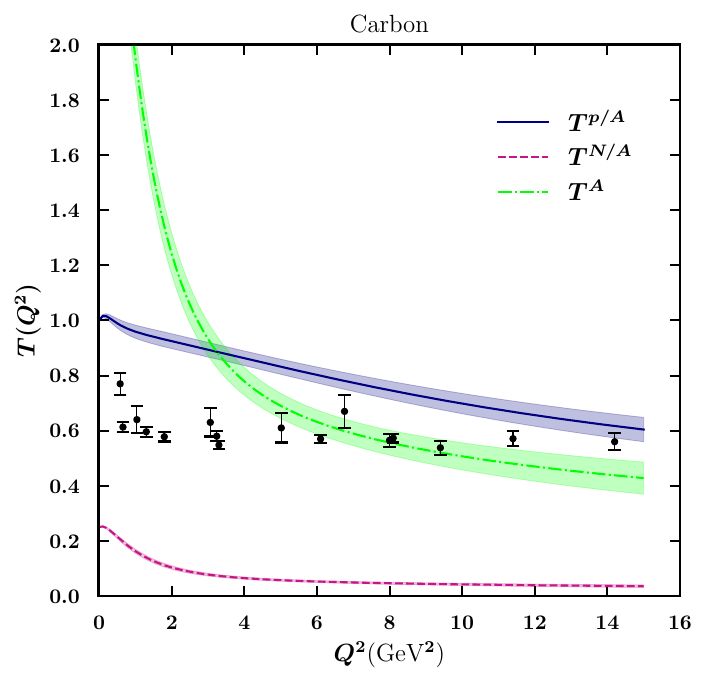}   
    \caption{Comparison of $ T^{p/A}(Q^2) $, $ T^{N/A}(Q^2) $, and $ T^A(Q^2) $ derived from the Dirac form factor $ F_1 $ as defined in Eqs.~(\ref{Eq12}) and~(\ref{Eq14}) and the experimental data of Refs.~\cite{Garino:1992ca,Makins:1994mm,ONeill:1994znv,Abbott:1997bc,Garrow:2001di,HallC:2020ijh}.}
\label{fig:TF}
\end{figure}

\begin{figure}[!htb]
    \centering
    \includegraphics[scale=0.73]{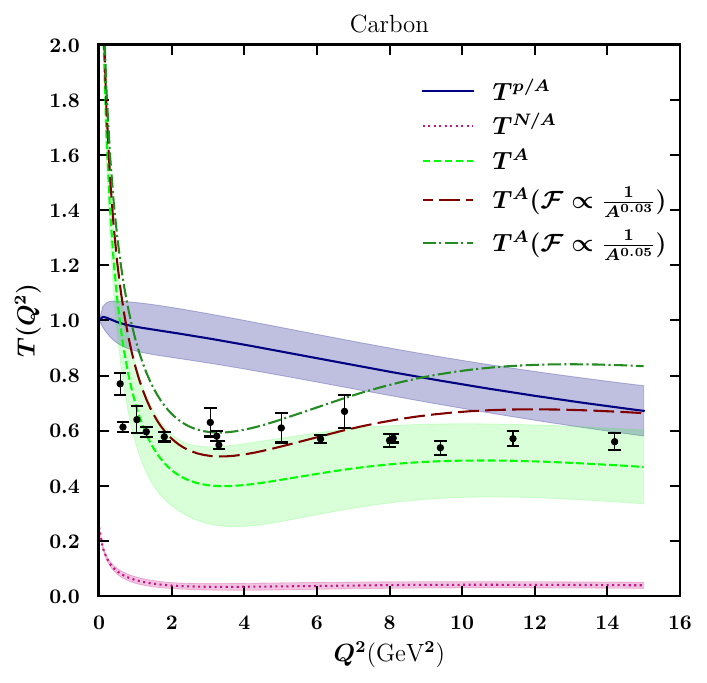}   
    \caption{Comparison of the results for $ T^{p/A}(Q^2) $, $ T^{N/A}(Q^2) $, and $ T^A(Q^2) $, based on the reduced cross-section $ \sigma_R $ as defined in Eq.~(\ref{Eq15}) and the experimental data of Refs.~\cite{Garino:1992ca,Makins:1994mm,ONeill:1994znv,Abbott:1997bc,Garrow:2001di,HallC:2020ijh}.}
    \label{fig:TSig}
\end{figure}

For the numerator $H^A$, we employ the two nGPD variants, namely $F^{p/A}$ and $ F^{N/A}$, using \texttt{nNNPDF} nPDFs~\cite{AbdulKhalek:2022fyi} at $\mu=2$ GeV. The denominator uses free proton GPDs from ``Set11" of Ref.~\cite{Hashamipour:2022noy}, based on \texttt{NNPDF} proton PDFs~\cite{NNPDF:2021njg} at the same scale. Figure~\ref{fig:TH} shows our calculations of $T(Q^2)$ defined in Eq.~(\ref{Eq11}) for Carbon nucleus using $H^{p/A}$ and $H^{N/A}$ nGPDs (labeled $T^{p/A}$ and $T^{N/A}$ respectively). The identical central values for both curves reflect Carbon's isoscalar nature ($Z = A-Z$)  for which the modifications of up quark distribution are the same as the ones of down quark distribution. As can be seen, the results obtained using Eq.~(\ref{Eq11}) disagree with experimental measurements~\cite{HallC:2020ijh} shown in Fig.~\ref{fig:TF}. They even show decreasing $T(Q^2)$ with increasing $Q^2$ which contradicts QCD predictions for color transparency. It should be noted that the error bands shown in Fig.~\ref{fig:TH}, as well as in all other figures throughout this paper, have been calculated using the standard Hessian method with $ \Delta \chi^2 = 1 $~\cite{Pumplin:2001ct}, taking into account the uncertainties from both PDFs and GPDs.

As previously discussed, we propose that the most accurate definition of nuclear transparency, $ T(Q^2) $, can be obtained by replacing GPDs in Eq.~(\ref{Eq11}) with the differential cross-section or reduced cross-section $ \sigma_R $. However, before doing so, it is instructive to first connect $ T(Q^2) $ to the Dirac form factor $ F_1(Q^2) $, given its established relationship with the GPD $ H $ as shown in Eq.~(\ref{Eq2}). This approach leads to the following expressions for $ T^{p/A}(Q^2) $ and $ T^{N/A}(Q^2) $:
\begin{equation}
 T^{p/A}(Q^2) = \frac{[F_1^{p/A}(Q^2)]^2}{[F_1^p(Q^2)]^2}\,, \qquad 
 T^{N/A}(Q^2) = \frac{[F_1^{N/A}(Q^2)]^2}{[F_1^p(Q^2)]^2}\,,
\label{Eq12}
\end{equation}
where the effective Dirac form factor for the nucleon in the nucleus, $ F_1^{N/A}(Q^2) $, is given by:
\begin{align}
F_1^{N/A}(Q^2) &= e_u \int_{0}^1 dx\, (H_v^u)^{N/A}(x,t) \nonumber \\ 
               &+ e_d \int_{0}^1 dx\, (H_v^d)^{N/A}(x,t)\,,
\label{Eq13}
\end{align}
with $ (H_v^q)^{N/A} $ obtained from Eq.~(\ref{Eq9}). Note that in contrast to Eq.~(\ref{Eq11}), which involves only the unweighted valence GPDs $ H_v^u $ and $ H_v^d $, the Dirac form factor $ F_1  $ automatically incorporates the electric charge factors $ e_u $ and $ e_d $ of the quarks. Therefore,  Eq.~(\ref{Eq12}) leads to a more physical description of the scattering process rather than Eq.~(\ref{Eq11}). Actually, Eq.~(\ref{Eq12}) is as an intermediate step toward the most direct definition of nuclear transparency in terms of the reduced cross section $ \sigma_R $, which we introduce later in the paper.

As mentioned before, Eq.~(\ref{Eq1}) does not describe coherent nuclear scattering; therefore, the GPD framework should be applied to bound nucleons. Actually, in quasielastic electron-nucleus scattering, the process probes the GPDs of bound nucleons rather than those of the entire nucleus. Consequently, the quasielastic nuclear cross section scales approximately with $A$. The nuclear transparency, defined as the ratio of the nuclear to the free-nucleon FFs, should thus also exhibit this $A$-scaling behavior. 
We therefore propose an alternative and more physically meaningful definition of the nuclear transparency, motivated by the underlying dynamics:
\begin{equation}
 T^{A}(Q^2) = \frac{A\,[F_1^{N/A}(Q^2)]^2}{[F_1^p(Q^2)]^2}\,.
\label{Eq14}
\end{equation}
This definition links nuclear transparency to a quantity that naturally incorporates the quark-charge weighting through the Dirac form factor $F_1(Q^2)$ while ensuring proper normalization at the nuclear level. Note that the $ A $ dependence originates from the incoherent sum over bound nucleons contributing to the quasielastic process.

Figure~\ref{fig:TF} presents the resulting calculations for $ T^{p/A}(Q^2) $, $ T^{N/A}(Q^2) $, and $ T^A(Q^2) $ based on the $ F_1 $ form factor in comparison with the experimental data of Refs.~\cite{Garino:1992ca,Makins:1994mm,ONeill:1994znv,Abbott:1997bc,Garrow:2001di,HallC:2020ijh} for Carbon nucleus. As shown, both $ T^{p/A}(Q^2) $ and $ T^{N/A}(Q^2) $ exhibit significant discrepancies with the experimental data~\cite{HallC:2020ijh}, while the definition based on $ T^A(Q^2) $ provides a much closer match to the observed trend.

We now repeat the calculations by defining $ T(Q^2) $ in terms of the ratio of reduced cross-sections, as follows:
\begin{align}
 T^{p/A}(Q^2) &= \frac{\sigma_R^{p/A}(Q^2)}{\sigma_R^p(Q^2)}\,, \qquad 
 T^{N/A}(Q^2) = \frac{\sigma_R^{N/A}(Q^2)}{\sigma_R^p(Q^2)}\,, \qquad \nonumber \\ 
 T^{A}(Q^2) &= \frac{A\, \sigma_R^{N/A}(Q^2)}{\sigma_R^p(Q^2)}\,,
\label{Eq15}
\end{align}
where $ \sigma_R $ expressed in terms of the Sachs electromagnetic FFs $ G_E $ and $ G_M $ as:
\begin{equation}
\label{Eq16}
\sigma_R = \epsilon G_E^2(Q^2) + \tau G_M^2(Q^2)\,.
\end{equation}
These form factors are themselves related to the Dirac and Pauli form factors, $ F_1 $ and $ F_2 $, as outlined in Eq.~(\ref{Eq3}). The dimensionless kinematic variables $ \epsilon $ and $ \tau $ are given by:
\begin{equation}
\label{Eq17}
\begin{split}
	\tau = \frac{Q^2}{4m^2}\,, \qquad
	\epsilon = \left[ 1 + 2 (1 + \tau) \tan^2 \left( \frac{\theta}{2} \right) \right]^{-1}\,,
\end{split}
\end{equation}
where $ m $ is the nucleon mass and $ \theta $ is the electron scattering angle.

According to the equations above, if nuclear transparency $ T(Q^2) $ is defined in terms of the reduced cross-section, then the contribution of the GPD $ E $ must also be taken into account, since the Pauli form factor $ F_2 $ is directly related to $ E $. To carry out the calculations, we also require point-by-point values of the electron scattering angle $ \theta $ in order to evaluate $ \epsilon $ in Eq.~(\ref{Eq17}). As $ \theta $ is experiment dependent, we simplify the calculation at this stage by assuming $ \tan^2 \left( \frac{\theta}{2} \right) = 1 $.

Another important consideration is the nuclear modification of the GPD $ E $. Unlike $ H $, for which nuclear modifications can be constrained from high-energy data, corresponding modifications for $ E $ are not currently available as mentioned before. However, we can at least incorporate isospin effects by defining $ (E_v^q)^{N/A} $ and $ (E_v^q)^A $ as follows:
\begin{align}
 (E_v^q)^{N/A}(x,t) &= \frac{Z}{A}(E_v^q)^{p/A}(x,t) + \frac{A-Z}{A}(E_v^q)^{n/A}(x,t)\,, \nonumber \\ 
 (E_v^q)^A(x,t) &= Z(E_v^q)^{p/A}(x,t) + (A-Z)(E_v^q)^{n/A}(x,t)\,,
\label{Eq20}
\end{align}
where $ (E_v^q)^{p/A} $ and $ (E_v^q)^{n/A} $ are taken to be the same as those for the free proton and neutron, respectively, in our calculations. Note that, according to our investigations, following scenario 2 mentioned before (universal nuclear modification) does not lead to significant changes in the final results.

Figure~\ref{fig:TSig} presents the results for $ T^{p/A}(Q^2) $, $ T^{N/A}(Q^2) $, and $ T^A(Q^2) $, this time calculated using the reduced cross-section $ \sigma_R $, in comparison with the experimental data of Refs.~\cite{Garino:1992ca,Makins:1994mm,ONeill:1994znv,Abbott:1997bc,Garrow:2001di,HallC:2020ijh} . As seen previously, both $ T^{p/A}(Q^2) $ and $ T^{N/A}(Q^2) $ show significant deviations from the experimental data. In contrast, the result for $ T^A(Q^2) $ exhibits much better agreement with the data presented in Fig.~2 of Ref.~\cite{HallC:2020ijh}. Indeed, the results in Fig.~\ref{fig:TSig} confirm that only the model based on $ T^A $ incorporates the correct $ A $ scaling and provides a realistic description of the data. The other definitions, $ T^{p/A} $ and $ T^{N/A} $, were intentionally presented as intermediate steps in order to demonstrate how neglecting the proper nuclear normalization leads to unphysical or $ A $-independent results. This step-by-step exploration is meant to motivate $ T^A $ as the physically consistent definition of nuclear transparency in the GPD framework.

It is important to note that the remaining discrepancy between our results for $ T^A(Q^2) $ and the experimental data may arise from several sources:

\begin{itemize}
\item Nuclear modifications of the GPD $ E $ (more precisely $e_v^q$) have not been included, as such modifications are currently unavailable. Although we accounted for the isospin effect, it has no impact in the case of a carbon nucleus, which contains an equal number of protons and neutrons. However, it should be noted that assuming identical nuclear modification factors for $e_v^q$ as those determined for the corresponding PDFs $q_v$ (scenario 2) does not change the final results significantly.

\item The profile functions $ \mathcal{F} $ have been treated as independent of the mass number $ A $, which may be an oversimplification. If this assumption is incorrect, $ \mathcal{F} $ should exhibit $ A $ dependence (more precisely, its individual parameters), and its functional form should ideally be determined through a global analysis of GPDs incorporating nuclear data. As an initial step, we introduce a simple $ A $-dependent modification to $ \mathcal{F} $ in Eq.~(\ref{Eq5}) and repeat the calculations to assess its impact on the results. Specifically, we consider $ \mathcal{F} \propto \frac{1}{A^\alpha} $, inspired by analogous analyses in the context of nPDFs~\cite{Hirai:2007sx}. So, the final form of the bound proton profile function is $\mathcal{F}^{p/A}(x)= \frac{1}{A^\alpha} f_v(x)$, where $f_v(x)$ is the free-proton profile function taken from Set11 of Ref.~\cite{Hashamipour:2022noy} for each quark flavor. The corresponding results are shown as the long-dashed and dot-dashed curves in Fig.~\ref{fig:TSig}, for $ \alpha = 0.03 $ and $ \alpha = 0.05 $, respectively. As can be seen, this modification leads to an enhancement in the result, particularly at higher values of $ Q^2 $. It should be noted that these values of $ \alpha $ are chosen nominally to explore the magnitude of the change induced by considering an overall $ A $-dependence in $ \mathcal{F} $. Using negative values for $ \alpha $ leads to a suppression of the final result. We emphasize that the true way to determine the $ A $ dependence of $ \mathcal{F} $ is to make its parameters explicitly $ A $-dependent and to extract their optimal values through a fit to relevant data. 

\item For the results shown in Fig.~\ref{fig:TSig} the scattering angle was assumed to be fixed ($ \theta = 90^\circ $). To accurately reflect the experimental data, the calculations were repeated using the exact value of $ \theta $  corresponding to each experimental data point.
The updated results are displayed in Fig.~\ref{fig:AngleStudy}, where the open circles represent the angle‑corrected theoretical predictions. The change is negligible at high momentum transfer ($ Q^{2} \gtrsim 5~\mathrm{GeV}^{2} $), but it becomes significant at lower $ Q^{2} $. In the interval $ 3 < Q^{2} < 4~\mathrm{GeV}^{2} $ the agreement with the data is modestly improved, whereas for $ Q^{2} < 2~\mathrm{GeV}^{2} $ the discrepancy grows. These observations  confirm again the need to determine the exact $ A $ dependence of the profile function $ \mathcal{F} $ through a dedicated fit to data. Note that individual parameters of $ \mathcal{F} $ may exhibit different $ A $-dependencies since they govern distinct regions of $ Q^{2} $. The results, in particular, suggest a significant modification in the parameter $ \alpha' $, which governs the behavior at low $ Q^2 $ and may reflect the limits of this formalism at low energy scales.
\begin{figure}[!htb]
    \centering
    \includegraphics[scale=0.73]{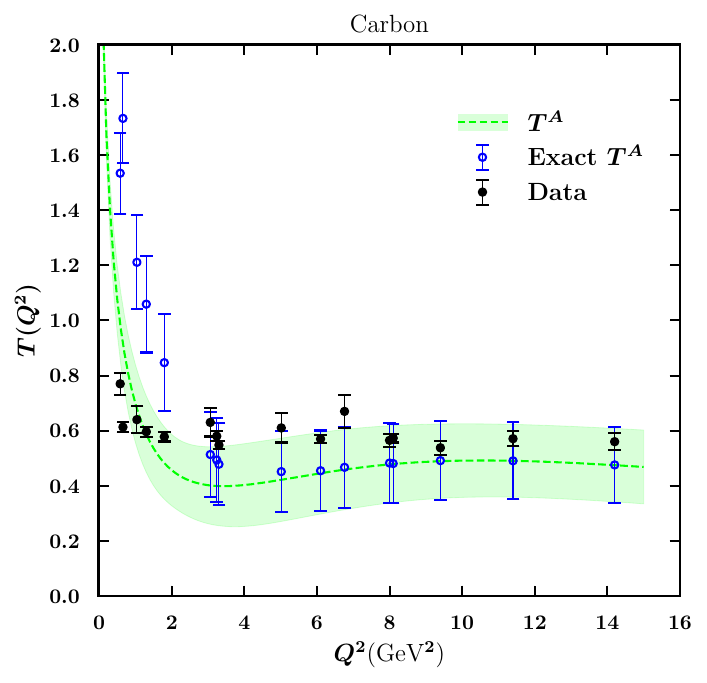}   
    \caption{Comparison of nuclear transparency calculations using an overall scattering angle $ \theta = 90^{\circ} $ (dashed line) versus using the exact scattering angles corresponding to each data point (open circles). The experimental data are taken from Refs.~\cite{Garino:1992ca,Makins:1994mm,ONeill:1994znv,Abbott:1997bc,Garrow:2001di,HallC:2020ijh}.}
    \label{fig:AngleStudy}
\end{figure}

\end{itemize}

%%%%%%%%%%%%%%%%%%%%%%%%%%%%%%	
	
	\textit{\textbf{\textcolor{violet}{Conclusions}}}~~
In this work, we have revisited the phenomenon of nuclear transparency in the context of CT using the framework of GPDs. By constructing nGPDs from known nPDFs, we have introduced a consistent approach to describing exclusive scattering processes involving nuclei. Our study focused on the role of CT in high-energy exclusive processes, where the propagation of compact hadronic states through nuclear matter should, in principle, lead to a characteristic rise in transparency with increasing $ Q^2 $. However, recent precision measurements of  nuclear transparency in carbon nucleus~\cite{HallC:2020ijh} indicated an unexpected $ Q^2 $-dependency that challenges this expectation.

To shed light on this issue, we calculated the nuclear transparency $ T(Q^2) $ for the carbon nucleus considering various definitions and compared our theoretical predictions with available experimental data. Our results confirm that the choice of definition for $ T(Q^2) $ remarkably affects the predicted behavior. Among the definitions considered, we find that the one formulated in terms of the reduced cross section and incorporating the correct $ A $ scaling provides the best agreement with the observed nuclear transparency up to $ Q^2 = 14.2 \, (\mathrm{GeV}/c)^2 $, as reported in recent measurements~\cite{HallC:2020ijh}. Additionally, we explored a simple $ A $-dependent modification to the profile function $ \mathcal{F} $ used in the GPD parametrization. While this modification does impact the magnitude and trend of $ T(Q^2) $, we emphasize that a more rigorous and reliable determination of nGPDs must be performed through a global analysis that incorporates nuclear data for different nuclei and kinematic conditions.

In conclusion, our findings highlight two key points: first, the definition of nuclear transparency should be carefully chosen to reflect the physical interpretation of medium modifications; second, achieving accurate and consistent results requires that the parameters of nGPDs be treated as nuclear dependent and fitted through global analyses. These insights open new windows towards future studies aiming to improve our understanding of CT and the internal structure of hadrons in the nuclear environment.

%%%%%%%%%%%%%%%%%%%%%%%%%%%%%%

	\textit{\textbf{\textcolor{violet}{Acknowledgements}}}~~
M. Goharipour thanks the Theoretical Physics Department at CERN for their kind hospitality during the period in which part of this work was carried out. F. Irani and K. Azizi are tankful to Iran National Science Foundation (INSF) for financial support provided for this research under grant No. 4033039.

	\onecolumngrid

	\twocolumngrid

%%%%%%%%%%%%%%%%%%%%%%%%%%%%%%%%%%%%%%%%%%%%%%%%%%%%%%%%%%

\bibliographystyle{apsrev4-1}
\bibliography{article} % References are stored in references.bib
	
	\onecolumngrid

\end{document}